# XPS Analysis of Surface Chemical Transformations in ZBLAN Glass under Thermal and Vibrational Stimuli

Ayush Subedi[1*], Anthony Torres[1], Jeff Ganley[2]

[1]*Materials, Science, Engineering, and Commercialization (MSEC), Texas State University, San Marcos, TX 78666, USA*
[2]*Air Force Research Labs, Space Vehicles Directorate, Kirtland Air Force Base, NM 87117, USA*
*Corresponding author ayush.rajsubedi@gmail.com

## Abstract

ZBLAN ($ZrF_4$-$BaF_2$-$LaF_3$-$AlF_3$-$NaF$) glass is highly sensitive to thermal and mechanical stimuli, yet the associated surface chemical changes remain poorly understood. X-ray Photoelectron Spectroscopy (XPS) measurements were performed on multiple ZBLAN samples representing distinct structural states: fully amorphous, incipiently crystalline, and highly crystalline, produced through thermal treatments at 250 °C and 350 °C and vibration-assisted processing at 400 °C under low (L2) and high (H5) vibration levels. High-resolution F 1s, Zr 3d, Hf 4f, Ba 3d, La 3d, and Na 1s spectra show progressive peak sharpening and intensity enhancement with increasing temperature and vibration, indicating reduced surface disorder and greater local structural ordering. The most pronounced changes occur under high vibration at 400 °C. No binding-energy shifts were detected, confirming that all elements retain their expected oxidation states and that the observed evolution reflects structural rather than chemical changes. These results provide direct evidence that thermomechanical input enhances surface ordering in ZBLAN and clarify its role in crystallization behavior relevant to infrared optical applications.

**Graphical Abstract**

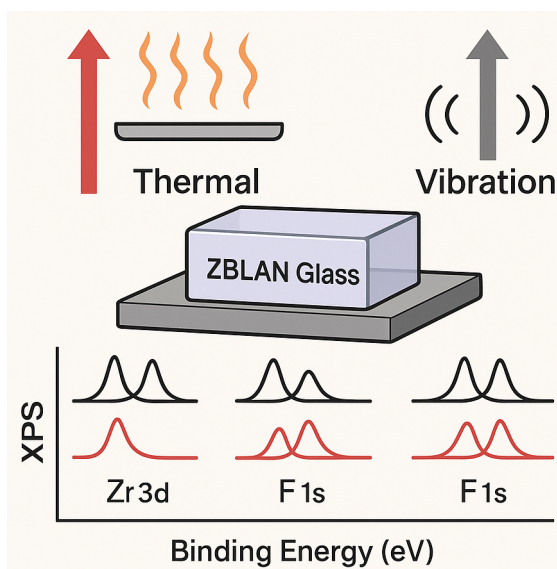

## 1. Introduction

ZBLAN glass ($ZrF_4$-$BaF_2$-$LaF_3$-$AlF_3$-$NaF$) is a widely utilized fluoride-based optical material valued for its broad infrared transparency and favorable optical performance [1]. Despite its technological significance, ZBLAN remains intrinsically metastable, and its structural integrity is highly sensitive to thermal and mechanical perturbations [2]. Understanding the mechanisms that govern its transition from an amorphous state to crystalline phases is therefore essential for optimizing processing strategies and improving long-term material stability.

In our previous work [2-3], we developed a vibration-assisted thermal-processing platform to systematically examine the crystallization behavior of ZBLAN under controlled thermal and mechanical excitation. Experimental investigations revealed that moderate vibrational input accelerates nucleation and decreases the crystallization onset temperature, whereas excessive vibration disrupts the sample-ampoule contact, promoting heterogeneous phase formation [2]. Complementary COMSOL Multiphysics simulations confirmed that these intermittent contact events significantly modify the effective heating rate and generate local temperature gradients, thereby shaping the observed crystallization pathways [3]. Together, these findings established that thermomechanical interactions fundamentally dictate the structural evolution of ZBLAN, influencing the boundary between amorphous and crystalline states.

Although morphological and bulk structural transitions under these conditions have been extensively characterized through optical microscopy, SEM, AFM, and EDS, the accompanying surface chemical dynamics remain largely unexplored. Surface composition and chemical state evolution are particularly critical in fluoride glasses, where fluorine depletion, and metal-fluoride bond rearrangement can substantially alter nucleation kinetics, diffusion behavior, and defect formation. A surface-sensitive analytical approach is therefore required to complement structural observations with direct chemical evidence.

The present work addresses this gap by employing X-ray Photoelectron Spectroscopy (XPS) to investigate the surface chemical transformations in ZBLAN exposed to varying thermal and vibrational environments. By analyzing binding-energy shifts and changes in elemental intensity ratios for Zr, Hf, Ba, La, Al, Na and F, we establish correlations between surface chemical states and the degree of structural ordering from fully amorphous to highly crystalline phases. These insights provide crucial evidence linking thermomechanical stimuli to compositional and bonding rearrangements, contributing to a more comprehensive understanding of ZBLAN's crystallization mechanisms and their implications for infrared optical applications.

## 2. Methodology

X-ray Photoelectron Spectroscopy (XPS) was used to investigate the surface chemical evolution of ZBLAN glass subjected to different thermal and vibrational treatments. Five representative samples were selected to capture progressive structural changes: an untreated amorphous reference (0V0C), a moderately heated but still amorphous sample (250 °C), a partially ordered sample near the onset of crystallization (350 °C), and two specimens treated at 400 °C under low (L2) and high (H5) vibration levels, as described in the previous study [2-3]. These samples span the transformation pathway from amorphous to increasingly ordered states and allow direct comparison with previously studied crystallization behavior.

All XPS measurements were performed using a Thermo Scientific Axia XPS system at the Shared Research Operations (SRO) Center, Texas State University, equipped with a monochromatic Al Kα X-ray source (1486.6 eV). Spectra were collected using a pass energy of 200 eV and an energy step size of 1.0 eV, with charge compensation applied during acquisition. Binding energies were referenced to the adventitious carbon C 1s peak at 284.8 eV to correct for surface charging effects. Prior to measurement, the sample surface was cleaned using a 3 keV monatomic ion beam with a 2.0 mm raster size to remove adsorbed contaminants while minimizing alteration of intrinsic surface chemistry. Given the ~5-8 nm information depth of XPS at this photon energy, the results reflect near-surface structural modifications rather than bulk crystallization.

High-resolution spectra were acquired for the F 1s, Zr 3d, Hf 4f, Ba 3d, La 3d, and Na 1s core levels, which provide sensitive indicators of bonding environments and coordination changes in fluorozirconate glasses. The Al 2p peak was not detected, consistent with its concentration being below the XPS detection limit for these operating conditions. Each core-level spectrum was examined for peak position, width, and intensity to evaluate chemical-state stability, near-surface structural ordering, and changes in elemental distribution induced by thermal and vibrational processing.

## 3. Results and Discussion

XPS measurements were conducted to evaluate the surface chemical composition of ZBLAN samples subjected to various thermal and vibrational treatments. Core-level spectra were collected for F 1s, Zr 3d, Hf 4f, Ba 3d, La 3d, and Na 1s across all conditions. All expected elements were detected except for Al 2p, whose absence is attributed to its concentration being below the XPS detection limit, consistent with earlier findings by Rizzato [4]. The complete set of elemental spectra is presented in Figure 1-6 and discussed below.

Figure 1 presents the F 1s core-level spectra for five representative samples, including the untreated reference (0V0C), thermally treated samples at 250 °C and 350 °C, and vibration-assisted samples processed at 400 °C under low (L2_400) and high (H5_400) vibration levels. The dominant F 1s peak appears near approximately 685 eV, characteristic of fluorine coordinated with metal cations such as Zr, Ba, La, and Al within the ZBLAN network [4]. The peak shapes were evaluated based on intensity, width, and symmetry, enabling qualitative comparison of fluorine bonding environments under different treatment conditions.

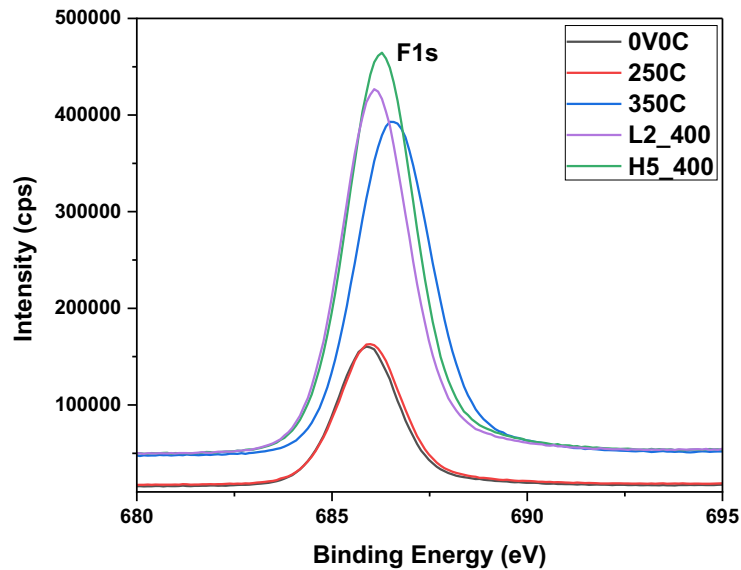

*Figure 1: F 1s peak spectra of ZBLAN glass showing fluorine bonding variations across amorphous to crystalline samples under different treatments.*

The untreated 0V0C sample exhibits a broad and relatively low-intensity F 1s peak, consistent with an amorphous surface containing heterogeneous and undercoordinated fluorine environments. Thermal treatment at 250 °C produces a modest increase in peak intensity and narrowing, suggesting partial relaxation and densification of the near-surface region. At 350 °C, the peak becomes sharper and more defined, suggesting the onset of short-range ordering and early-stage structural reorganization associated with initial crystallization processes.

The most substantial spectral evolution occurs in the vibration-assisted samples treated at 400 °C. The L2_400 sample shows a more pronounced and symmetric peak compared to the purely thermally treated samples, reflecting enhanced atomic mobility and redistribution of fluorine into more uniform metal-fluoride bonding environments. The H5_400 sample exhibits the highest peak intensity and sharpest profile among all conditions, reflecting a substantial reduction in surface disorder and the formation of highly ordered fluorine coordination environments.

Importantly, no binding-energy shifts were observed in any of the samples, confirming that fluorine remains in the same chemical state across all treatment conditions. The observed spectral changes therefore arise from structural rather than compositional modifications. The progressive narrowing and intensification of the F 1s peak demonstrate increasing uniformity of the surface fluoride environment under higher thermal and vibrational inputs.

Figure 2 shows the Zr 3d spectra for the five samples investigated. The Zr 3d region typically consists of a $3d_{5/2}$ & $3d_{3/2}$ spin-orbit doublet separated by ~2.4 eV [4-6], although in this study only the Zr $3d_{5/2}$ component is clearly observed near ~179.5 eV, while the $3d_{3/2}$ peak remains unresolved.

This is consistent with XPS measurements of low-intensity Zr signals, where the weaker $3d_{3/2}$ line can be obscured by background noise or overlap with adjacent features.

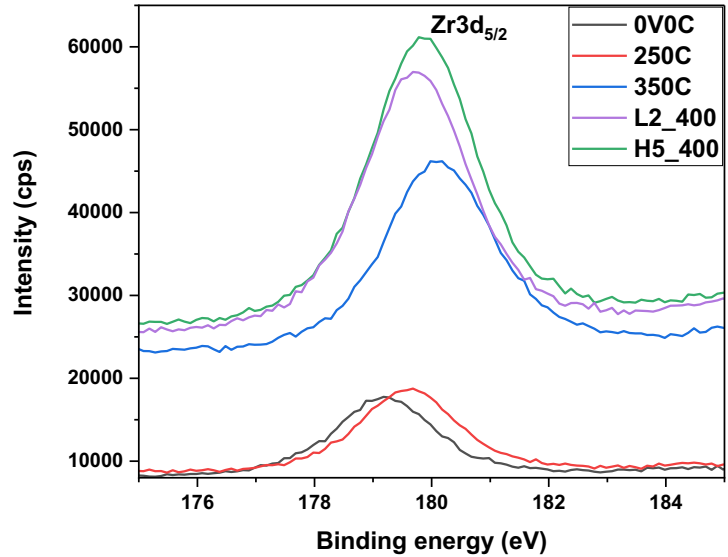

Figure 2: Zr 3d peak spectra of ZBLAN glass showing zirconium bonding variations across amorphous to crystalline samples under different treatments.

The untreated 0V0C sample exhibits a weak and broad Zr $3d_{5/2}$ peak, indicating a disordered near-surface environment with limited zirconium signal. With increasing thermal treatment temperature, the peak intensity increases and the width narrows, reflecting improved surface ordering. The most pronounced enhancement occurs in the samples treated at 400 °C under vibration (L2_400 and H5_400), with H5_400 showing the strongest and sharpest Zr $3d_{5/2}$ peak. This suggests that elevated temperature combined with mechanical vibration increases atomic mobility and promotes a more uniform zirconium bonding environment at the surface.

No measurable shift in the Zr $3d_{5/2}$ binding energy is observed across the samples, confirming that zirconium remains in the $Zr^{4+}$ state and that the spectral changes arise from structural reorganization rather than chemical alteration.

Figure 3 presents the Hf 4f spectra of ZBLAN samples. The Hf 4f spectrum typically consists of a spin orbit doublet corresponding to Hf $4f_{7/2}$ and Hf $4f_{5/2}$ with separation of approximately 1.71 eV [5]. The Hf $4f_{7/2}$ peak appears near ~20 eV, while the weaker $4f_{5/2}$ component remains unresolved due to its lower intrinsic intensity and partial overlap with the surrounding spectral background, like the behavior observed in the Zr 3d spectra.

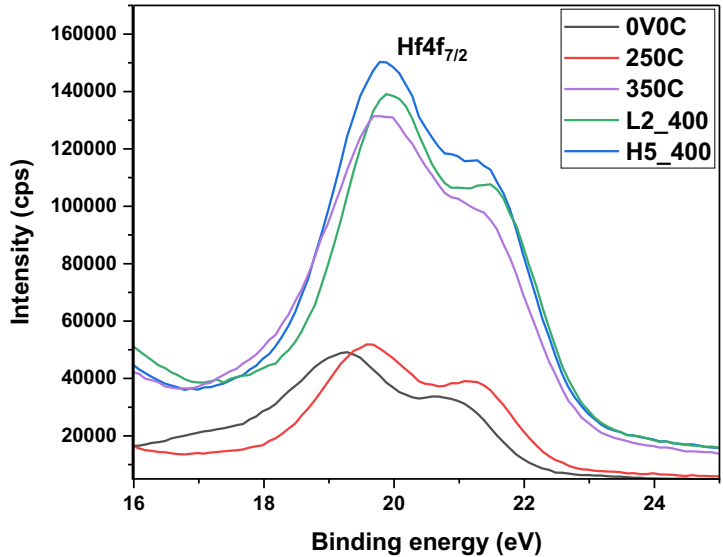

*Figure 3: Hf 4f peak spectra of ZBLAN glass showing hafnium bonding variations across amorphous to crystalline samples under different treatments.*

The untreated 0V0C sample exhibits a broad and low-intensity Hf $4f_{7/2}$ peak, indicative of a disordered near-surface environment and limited hafnium signal. With increasing thermal treatment temperature, the peak becomes progressively more intense and better defined, with a noticeable improvement at 350 °C. Further enhancement is observed in the vibration-assisted samples processed at 400 °C. Both L2_400 and H5_400 display sharper, higher-intensity peaks, with the H5_400 condition showing the strongest and most well-defined signal. These changes reflect increased uniformity in the local hafnium coordination environment and improved near-surface structural organization under combined thermal and vibrational input.

No measurable binding-energy shift is observed across the samples, confirming that hafnium remains in the $Hf^{4+}$ oxidation state under all processing conditions. The spectral evolution therefore reflects structural reorganization rather than chemical modification of hafnium species.

Figure 4 presents the Ba 3d spectra for the five ZBLAN samples. The Ba 3d region consists of a well-defined spin-orbit doublet, with the Ba $3d_{5/2}$ and Ba $3d_{3/2}$ peaks typically located near ~781 eV and ~796 eV, respectively, characteristic of $Ba^{2+}$ in barium fluoride [5]. Both components of the doublet are visible across all samples.

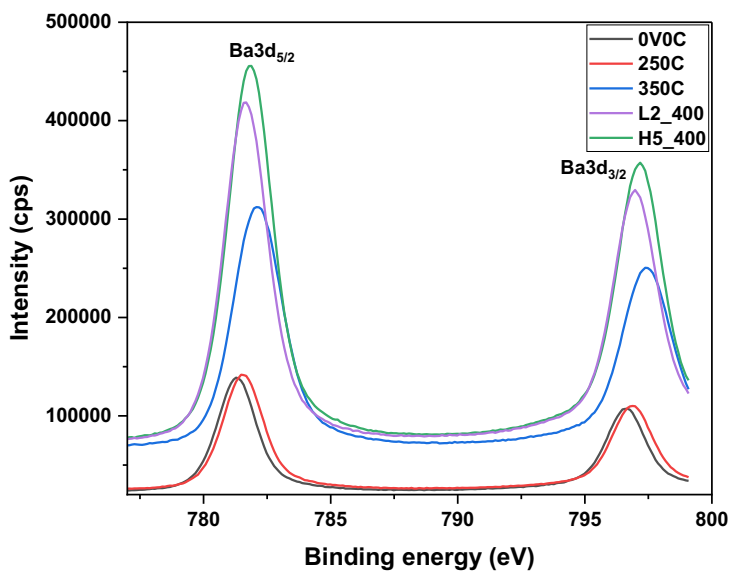

*Figure 4: Ba 3d peak spectra of ZBLAN glass showing barium bonding variations across amorphous to crystalline samples under different treatments.*

In the untreated 0V0C sample, the Ba 3d peaks are relatively broad and of lower intensity, reflecting a disordered amorphous surface environment. No noticeable change is observed at 250 °C; however, upon heating to 350 °C the peaks become sharper and more intense, indicating the onset of near-surface structural ordering. This trend becomes more pronounced in the samples treated at 400 °C under vibration. The L2_400 sample exhibits narrower and more intense Ba 3d peaks than the thermally treated samples, while the H5_400 condition produces the strongest and most well-resolved doublet among all measurements.

The consistent binding energies across all treatment conditions confirm that barium remains in the $Ba^{2+}$ oxidation state. The observed variations in peak intensity and width therefore reflect changes in structural order rather than changes in chemical state. The progressive sharpening and intensification of the Ba 3d peaks with increasing thermal and vibrational input indicate improved uniformity in the local barium-fluoride environment at the surface.

Figure 5 presents the La 3d spectra for the five ZBLAN samples. The La 3d region is characterized by a well-separated $3d_{5/2}$-$3d_{3/2}$ spin-orbit doublet, with each component further split into multiple features due to multiple interactions typical of $La^{3+}$ in fluoride environments [4-6]. In the present samples, two La $3d_{5/2}$ features are clearly visible in the range of ~834-835 eV. This double-feature appearance results from the well-known multiplet splitting of the $La^{3+}$ $3d_{5/2}$ transition, which occurs due to strong final-state interactions between the 3d core hole and the 4f electrons in lanthanum compounds, including lanthanum fluorides.

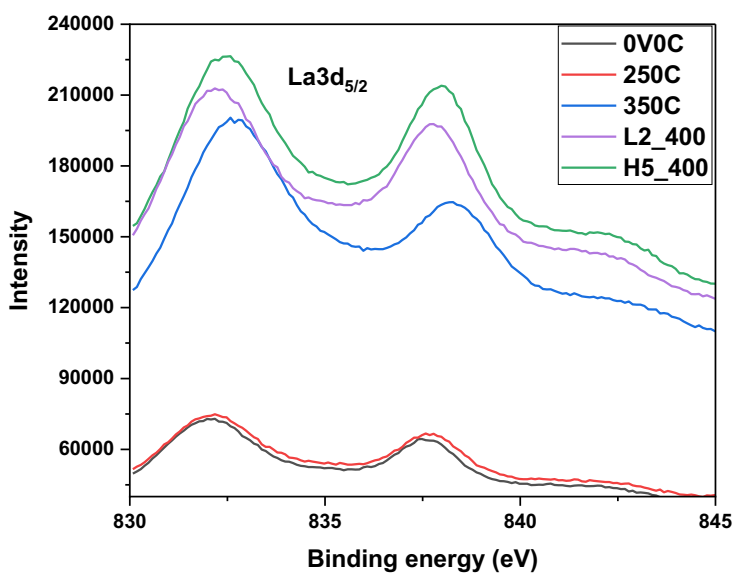

*Figure 5: La 3d peak spectra of ZBLAN glass showing lanthanum bonding variations across amorphous to crystalline samples under different treatments.*

In the untreated 0V0C sample, the La $3d_{5/2}$ peaks appear broad and weakly defined, consistent with a disordered amorphous surface layer. The sample treated at 250 °C shows similar peak shape and intensity, indicating minimal structural change at this temperature. In contrast, the 350 °C sample displays noticeably sharper and more intense La $3d_{5/2}$ features, signifying the onset of near-surface structural ordering. The effect becomes more pronounced in the vibration-assisted samples processed at 400 °C. The L2_400 sample shows improved peak definition and higher intensity relative to the purely thermally treated samples, while the H5_400 condition yields the strongest and most well-resolved La $3d_{5/2}$ peaks, reflecting the highest degree of ordering at the surface.

The La 3d binding energy positions remain consistent across all processing conditions, confirming that lanthanum maintains its $La^{3+}$ oxidation state throughout the treatments. The variations in peak intensity and sharpness therefore arise from structural reorganization rather than changes in chemical state. The progressive enhancement in spectral definition with increasing temperature and vibration indicates increasing uniformity in the local lanthanum-fluoride environment at the surface.

Figure 6 shows the Na 1s spectra for the five ZBLAN samples. The Na 1s peak, typically located in the 1071-1073 eV range, corresponds to $Na^+$ in sodium fluoride-type environments [4-6]. A single Na 1s feature is observed across all samples.

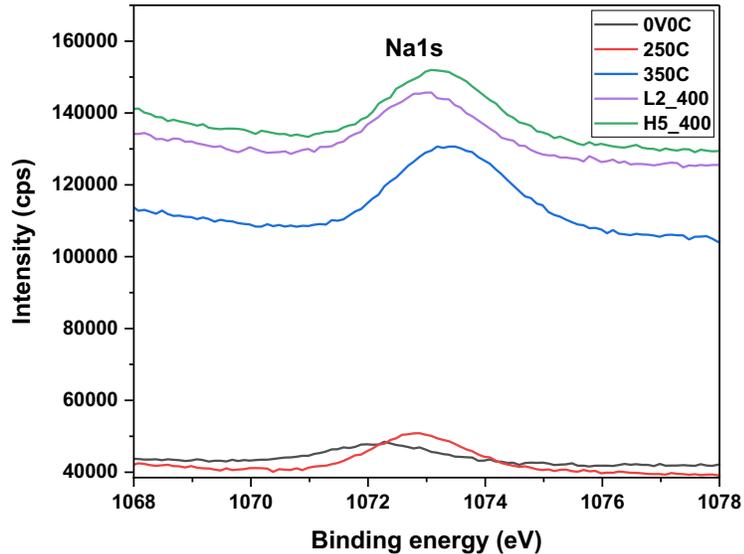

*Figure 6: Na 1s peak spectra of ZBLAN glass showing sodium bonding variations across amorphous to crystalline samples under different treatments.*

In the untreated 0V0C sample, the Na 1s peak is broad and of very low intensity, indicating a disordered near-surface sodium environment. The sample treated at 250 °C exhibits only a slight increase in peak intensity, with minimal change in overall shape, suggesting limited structural modification at this temperature. A more defined Na 1s signal emerges at 350 °C, consistent with the onset of local ordering within the glass.

The most substantial spectral evolution occurs in the samples processed at 400 °C under vibration. The L2_400 sample displays a notably stronger and better-resolved Na 1s peak, while the H5_400 condition produces the most intense and well-defined signal among all samples. This progressive enhancement indicates increased uniformity in the sodium environment at the surface, likely facilitated by higher atomic mobility and structural reorganization under combined thermal and vibrational input.

The Na 1s binding energy remains constant across all conditions, confirming that sodium persists in the $Na^+$ oxidation state. The observed differences in peak intensity and sharpness therefore reflect structural, rather than chemical, changes in the local sodium-fluoride environment at the surface.

### 4. Conclusion

This study systematically examined the surface chemical evolution of ZBLAN glass subjected to controlled thermal and vibrational treatments using X-ray Photoelectron Spectroscopy (XPS). Across all measured core-level spectra, including F 1s, Zr 3d, Hf 4f, Ba 3d, La 3d, and Na 1s, a consistent trend was observed: progressive sharpening and intensification of peaks with increasing temperature and, more prominently, under vibration-assisted conditions.

Thermal treatment up to 350 °C resulted in moderate peak narrowing and intensity enhancement, indicating the onset of surface structural ordering while preserving the original chemical states of the constituent elements. The most pronounced changes occurred in samples treated at 400 °C with mechanical vibration. Both low- and high-level vibration increased spectral definition, with the H5_400 condition producing the sharpest and most intense peaks across all elements. These changes reflect improved uniformity in local metal-fluoride environments and enhanced near-surface organization, consistent with early-stage crystallization or nanoscale ordering at the glass surface.

No binding-energy shifts were detected for any element, confirming that $Zr^{4+}$, $Hf^{4+}$, $Ba^{2+}$, $La^{3+}$ and $Na^+$, as well as fluorine, remained chemically unchanged throughout the treatments. The observed spectral evolution therefore arises from structural rather than compositional modifications. Together, the results show that the combination of elevated temperature and mechanical vibration significantly accelerates surface ordering in ZBLAN, promoting a more uniform and crystalline-like arrangement of metal-fluoride units.

These findings clarify the link between thermomechanical processing and surface chemical organization in fluoride glasses. Understanding how vibration enhances atomic mobility and restructures the near-surface region provides useful guidance for controlling crystallization behavior and improving the surface stability of ZBLAN in infrared photonic applications.